\documentstyle[aps,preprint,prl,tighten]{revtex}
\begin{document}
\title{The Smallest Molecular Switch}
\author{Eldon G. Emberly$^1$ and George Kirczenow$^2$}
\address{$^1$Center for Physics and Biology, Rockefeller University,
New York, NY 10028 \linebreak
$^2$Physics Department, Simon Fraser University, Burnaby, BC V5A 1S6}
\date{\today}
\maketitle
\begin{abstract}
{\em Ab-initio} total energy calculations reveal
benzene-dithiolate (BDT) molecules on a gold surface, contacted by
a monoatomic gold STM tip to have two classes of low energy
conformations with differing symmetries. Lateral motion of the tip
or excitation of the molecule cause it to change from one
conformation class to the other and to switch between a strongly
and a weakly conducting state. Thus, surprisingly, despite their
apparent simplicity these Au/BDT/Au nanowires are shown to be
electrically bi-stable switches, the smallest two-terminal
molecular switches to date. Experiments with a conventional or novel self-assembled STM are 
proposed to test these predictions.

\end{abstract}
%
%85.65.+h %Molecular electronic devices
%73.63.-b %Electronic transport in nanoscale materials and structures
%82.37.Gk %STM and AFM manipulations of a single molecule
%
%\begin{multicols}{2}

Electronic devices that switch between high and low resistance states  are at
the heart of the modern information technology. As miniaturization of this
technology continues to progress the long-standing fundamental problem of
identifying and understanding the smallest physical systems that are capable
of switching behavior is attracting growing interest
\cite{Reedsw,Collier,Donhauser,Sem,Swtheory,Damle1,Bratkovsky,Korswitch}.
Recently it has been discovered experimentally that some molecular wires
(i.e., {\em single molecules} carrying an electric current between a pair of
metal nano-electrodes) can exhibit electrical bistability and switch between
strongly and weakly conducting states, either spontaneously or in response to
a change in the applied bias voltage\cite{Reedsw,Collier,Donhauser}. It has
been suggested that this intriguing behavior may be due to charging of the
molecule and/or changes of the molecular geometry
(conformation)\cite{Reedsw,Collier,Donhauser,Sem}, however the complexity of
the experimental systems has so far prevented the development of a
quantitative explanation. On the theoretical side, the possibility of making
molecular wire switches by introducing a third (gate) electrode into the
system has been explored\cite{Swtheory,Damle1,Bratkovsky}, as has the
possibility of designing two-terminal molecular wires that switch due to
electric field-induced conformational changes\cite{Korswitch}. In this Letter
we demonstrate theoretically that  a much smaller and simpler two-terminal
molecular wire can exhibit bi-stability and switching than has been
thought possible until now, and present a realistic theory of its behavior.  We
consider a 1,4 benzene-dithiolate (BDT) molecular wire with one sulfur end
group bonded to a gold substrate and the other to a monoatomic gold scanning
tunneling microscope (STM) tip as depicted in Fig.1. Our {\em ab-initio}
total energy calculations\cite{Gaussian} demonstrate that this system has
low-energy conformations of two distinct types that have different
symmetries. Flipping between these conformations is predicted to occur in
response to lateral motion of the STM tip, and also, for some positions of
the STM tip, in response to excitation of the wire by a current pulse or
spontaneously at finite temperatures. The symmetry change when the molecular
wire flips results in a large change in its current-voltage characteristic.
Thus its electrical conductance exhibits bi-stability and switching.
Theoretical work\cite{Emberly98,Hall,DiVentra,Damle,Korn,Emberly} stimulated
by a pioneering molecular wire experiment\cite{Reed1} has elucidated various
aspects of electron transport through Au/BDT/Au wires. However the
possibility that wires of this type may be bi-stable or capable of switching
is far from obvious {\em a priori} and has not been investigated until now
\cite{expl}. Thus as well as identifying the smallest two-terminal molecular
switch to date and shedding new light on the mechanisms of molecular
bistability and switching, the present work reveals an unexpected new
dimension of the physics of Au/BDT/Au molecular wires, one of the most
important paradigms of molecular electronics. The molecular switch that we
describe should be amenable to experimental study with presently available
techniques. Thus our findings also raise the prospect of bridging the gap
that has persisted in this field between theory and experiment since
molecular switching was first observed.

In our {\em ab-initio} calculations of the energetics of Au/BDT/Au
wires\cite{Gaussian} the STM tip was represented by a tetrahedron of Au atoms
and the Au (111) substrate by a cluster of three Au atoms as shown in Fig.1.
We searched for low-energy molecular wire conformations holding the positions
of the Au atoms fixed and keeping one of the S atoms of the BDT over the
hollow site between the three Au substrate atoms since it is believed that
organic thiol molecules bond to Au (111) surfaces via a sulfur atom at this
location\cite{hollow}. All other coordinates of the atoms of the wire were
allowed to vary freely. The lowest energy conformation of the molecular wire
when the Au tip atom is directly over the hollow bonding site of the
substrate and 9.9\AA\ from the substrate is depicted in Fig.1(a). The
molecule orients itself so that the  Au tip atom is approximately coplanar
with the benzene ring. We will refer to this as an {\em edge} conformation of
the wire since the Au tip atom faces the edge of the benzene ring.  If the Au
tip is moved further from the substrate, so that it becomes {\em
geometrically} possible for the molecule to stand perpendicularly to the
substrate with the terminal Au atom of the STM tip directly over the upper S
atom, we find this upright geometry to be unstable energetically: As long the
tip atom is close enough to the molecule for a chemical bond to form between
the Au and S atoms, the molecule relaxes to a tilted position. We note that
such non-linear bonding geometries of Au, S and C atoms have also been found
in recent {\em ab-initio} simulations of monoatomic Au wires bonded to other
organic thiolate molecules
\cite{Kr}.

When the Au tip is moved laterally away from the position over the hollow
substrate bonding site, we find the molecular wire's ground state
conformation to change dramatically.  This is illustrated in Fig.1(b) where
the tip has been displaced laterally by $3$\AA in the
$x$ direction from its ($x=0$) position in Fig.1(a). Interestingly, the
molecule adopts an orientation in which it tilts as far from the normal to the
substrate as it can while maintaining the chemical bond between the Au tip
atom and the upper S atom of the molecule. Most importantly, however, as the
lateral tip displacement increases from $x \sim 2$\AA\ to
$\sim 3$\AA\ the molecule rotates about its S-S axis from its  {\em edge}
conformation through an angle of
$\sim \pi/2$ to an orientation in which the Au tip atom is over the flat face
of the benzene ring, as in Fig.1(b). We shall refer to this as a {\em face}
conformation.  Thus the ground state conformation of the molecular wire
switches from {\em edge} to {\em face} as the STM tip moves laterally away
from the location where the molecule bonds to the substrate. If the tip is
moved instead in the opposite (negative $x$) direction from its position in
Fig.1(a) the ground state conformation of the molecular wire again switches
from {\em edge} to a strongly tilted {\em face} geometry although the
substrate and tip both have a pronounced left-right asymmetry. For the
$x=3$\AA\ position of the tip in Fig.1(b) we find the molecular wire to also
have a metastable conformation that corresponds to a {\em local} energy
minimum. This is the upright {\em edge} conformation in Fig.1(c) whose energy
is 0.11eV above the ground state {\em face} geometry of Fig.1(b).

We now examine the implications of the conformational  switching and
bi-stability described above for electrical conduction through the molecular
wire. In recent years much progress has been made developing theories of
electron transport through molecules
\cite{Emberly98,Hall,DiVentra,Damle,Korn,Emberly,Datta97,Theory,Kushmerick,Review}.
An important conclusion has been that the current at low bias is carried by
molecular $\pi$ orbitals. The overlap between the $\pi$ orbitals and the
states of the contacts is sensitive to the orientation of the molecule
relative to the contacts, which implies a strong orientation-dependence of
the molecular wire's conductance\cite{Korn}. Such overlap effects have been
found in semi-empirical\cite{Emberly98,Emberly} and density functional
\cite{DiVentra} transport calculations. Thus it is reasonable to expect them
to result in a significant change in conductance when an Au/BDT/Au wire
switches between an {\em edge} and a {\em face} conformation, and our
calculations show this to be the case. Since semi-empirical calculations have
been successful in explaining the experimental current-voltage
characteristics of a variety molecular wires consisting of organic thiol
molecules bonded to gold electrodes\cite{Emberly,Datta97,Kushmerick} we
adopted this approach here\cite{calc}.  In Fig.~\ref{fig2} we show the
calculated differential conductance\cite{calc} of the molecular wire in its
{\em ground state} conformation for a sequence of positions of the STM tip
along a linear trajectory over the molecule that passes through the locations
that the tip occupies in Fig.1. (Fig.1(a) and (b) correspond to
$x=0$\AA\ and
$3$\AA\  on the trajectory, respectively.)      When the tip is furthest from
the center where the molecule bonds to the substrate (i.e. for
$x=-4 $\AA\  and $4$\AA\ in Fig.~\ref{fig2}) the molecule in its ground state
is in the {\it face} conformation and is highly conducting at a source-drain
bias around $1.5$ Volts. When the tip moves towards the center ($x=-2, 0, 2
$\AA\  in Fig.~\ref{fig2}), the molecule's ground state switches to the {\it
edge} configuration which is seen to be much less conducting in the same
range of bias.  Our calculations of the $dI/dV$ characteristics of the
molecular wire in a variety of  {\em face} and {\em edge} conformations that
have energies higher than the ground state (including the metastable {\em
edge} conformation in Fig.1(c)) yielded very similar results to those in
Fig.~\ref{fig2}. I.e., All {\em face} conformations of the molecular wire
were found to be highly conducting at the first conductance peak near $1.5$ V
while all {\it edge} conformations are weakly conducting there. Thus whenever
the molecule is made to flip from a {\em face} conformation to an {\it edge}
conformation, either by displacing the STM tip laterally or by exciting the
molecule thermally or by a current pulse, the molecule is predicted to switch
from a highly conducting to a weakly conducting state, and vice versa.

This large difference in conductance between the {\it edge} and {\em face}
conformations can be understood within Landauer theory \cite{Landauer} by
considering the transmission probabilities $T$  for electrons to scatter
through the molecular wire, taking account of $\pi$ orientational
effects\cite{Korn}. Representative results\cite{calc} are shown in
Fig.~\ref{fig3} for several {\it face} ( Fig.~\ref{fig3}(a)) and {\it edge}
(Fig.~\ref{fig3}(b)) conformations.   The Fermi energy of gold in our
semi-empirical model is near $-10$ eV \cite{calc}. The transmission peaks
immediately below the Fermi energy can be attributed to the highest occupied
molecular orbitals (HOMO)s of the BDT and those above to the lowest
unoccupied molecular orbital (LUMO). The Fermi energy lies nearest the
HOMO\cite{DFT}, so the onset of conductance is due to electron transmission
through  the HOMO.  In the {\it face} conformations there is a strong overlap
between the first HOMO molecular $\pi$ orbital and the atomic orbitals on the
Au tip atom. This results in the strong transmission due to the first HOMO
below the Fermi energy (beginning near $-10.5$ eV) in Fig.~\ref{fig3}(a) and
in the strong conductance peak seen for the {\it face} conformations in
Fig.~\ref{fig2}. However, in the {\em edge} conformations the molecular $\pi$
orbitals are oriented differently and their overlap with the orbitals on the
Au tip atom is weaker. This results in the weaker transmission through the
first HOMO below the Fermi energy in Fig.~\ref{fig3}(b) and the lower
conductance of the {\em edge} conformations in Fig.~\ref{fig2}.

Thus we arrive at the unexpected prediction that a molecule as simple  as BDT
can be made to switch through its interaction with a suitable STM tip.  The ON
state  corresponds to the molecule oriented in such a way that its ring faces
the tip, whereas in the OFF state the edge of the ring faces the tip. The
switching can be induced by passing the tip over the molecule (the transition
between Fig.1(a) and Fig.1(b))  or by exciting the molecule to a different
conformation for a fixed tip position (the transition between Fig.1(b) and
Fig.1(c)).

We predict an unusual and striking experimental signature of switching
induced by the motion of the tip:  The conductance should be {\em low at the
center} of the STM image of the molecule where the {\em edge} conformation
is stable and {\em high} when the  tip moves {\em away} from the center of
the image and the {\em face} conformation becomes stable. Experimental
observation of switching induced by passing the tip over the molecule may be
facilitated at low temperatures where thermal excitation of the higher energy
conformations is minimal. [For example, we find the highly conducting excited
{\em face} conformation obtained by rotating the molecule from the (weakly
conducting) ground state {\em edge} conformation  in Fig.1(a) through $\pi/2$
about the S-S axis corresponds to an energy saddle point $\sim kT_{room}$
above the ground state; thus this excited state will be populated
significantly at room temperature.] Hysteresis may occur in the STM image of
the molecule at low temperatures, as is evident from the greater similarity
of the $x=0$\AA\ ground state in Fig.1(a) to the $x=3$\AA\ {\em excited}
state in Fig.1(c) than to the ground state in Fig.1(b). This also suggests
interesting experiments that may combine scanning by the tip and activation
of switching by current pulses.

Experimental studies may employ a conventional STM to contact a molecule
adsorbed on a metal substrate\cite{expl}. An alternative may be to {\em
self-assemble} a monoatomic STM tip  for the top probe. Realizing such a
system would rely on growing the top electrode epitaxially on a heterogeneous
self-assembled monolayer (SAM) of insulating and conducting organic molecules,
choosing a conducting molecule that is shorter than the insulating one. Such
a SAM would have a ``divot'' at the site of each conducting molecule.  It has
been suggested\cite{Bratkovsky} that when the top layer is grown, a metal
atom that bonds chemically to the conducting molecule  should occupy the
divot forming a tip. Our calculations indicate that the top electrode may sit
quite high over the insulating part of the SAM (we considered gold contacts
with pentanethiol as the insulator and BDT as the conductor). Thus there is
room for the formation of a monoatomic tip contacting the conducting
molecule. In such systems the switching rate between the ON ({\em face}) and
OFF ({\em edge}) state for a fixed tip position would depend on steric
constraints  imposed by the local environment of the molecule; analogous
effects have been observed recently in STM experiments on more complicated
(three-ring) switching molecules in an insulating host
matrix\cite{Donhauser}. For an insulating host SAM such as an alkanethiol
that does {\em not} bond chemically to the  upper metal contact, one may also
consider the possibility of sliding the top metal electrode laterally a few
Angstroms (and the self-assembled tip with it) relative to the substrate and
SAM by applying suitable mechanical or electrostatic forces to the metal
contacts, thus realizing a {\em mobile} self-assembled STM tip that could
probe switching of the molecule induced by motion of the tip.

In conclusion, we have shown that one of the simplest and most studied
molecular  wires, surprisingly, becomes a bistable molecular switch if one of
its two metal contacts is a monoatomic scanning tunneling microscope tip.
This is the smallest two-terminal molecular switch to date. The switching
mechanism that we have introduced here relies on the coupling between the
molecule and contacts and thus should be broadly applicable. We have proposed
experiments with a conventional or novel {\em self-assembled} STM to test our
predictions, and hope that the ideas put  forward here will facilitate
bridging the gap between theory and molecular switching experiments.

This work was supported by NSERC and the Canadian Institute for 
Advanced Research.

%\end{multicols}

\begin{figure}[!t]
\caption{Color. Relaxed molecular conformations. (a) Ground state ({\it
edge}) conformation for STM tip at $x =
0 $\AA, over hollow substrate bonding site (b) Ground state ({\it 
face}) conformation
for $x=3$ \AA. (c) Metastable {\it edge} conformation for $x=3$ \AA.
\label{fig1}}
\end{figure}

\begin{figure}[!t]
\caption{Color. Calculated differential conductance for ground state BDT
conformations at several STM tip positions. At $1.5$ volts
the molecule is in the ``ON'' state when the tip is at $x=-4$ and $4$ 
\AA and ``OFF'' at
other $x$ values. \label{fig2}}
\end{figure}

\begin{figure}[!t]
\caption{Color. Calculated transmission probabilities $T$ vs.
electron energy $E$ for different tip positions. (a)
Transmission for ground state {\it face} conformations. (b)
Transmission for some {\it edge} conformations (the
metastable edge conformation is the excited state
at $x=3$\AA\ in Fig.1(c)).  The Fermi energy is $-10$ eV. Resonances
due to the HOMO are below the Fermi energy. \label{fig3}}
\end{figure}


\begin{references}
\bibitem{Gaussian} We employed the Gaussian 98 numerical 
implementation of density
functional theory with the Lanl2DZ basis set
and the B3PW91 exchange-correlation energy functional.
\bibitem{expl}Experiments on Au/BDT/Au reported in Ref.\onlinecite{Reed1}
employed a break-junction that did not allow lateral scanning,
may have involved transport through more than one molecule as
was suggested in Ref.\onlinecite{Emberly}, and were at
room temperature. Thus they were not
suitable for detecting the switching behavior that we predict
here.
\bibitem{calc} The system was taken
to consist of face centered cubic (100) semi-infinite gold leads
with exposed (111) faces to which the STM tip and
relaxed molecule were attached. The Landauer transmission
probabilities \cite{Landauer}
and differential conductance of the molecular wires
were then calculated using the semi-empirical tight-binding scattering
formalism described in Ref. \onlinecite{Emberly} and references therein.
\bibitem{DFT} This is consistent with density functional calculations
of electronic structure of Au/BDT/Au reported in Ref.\onlinecite{Damle1}.
\end{references}
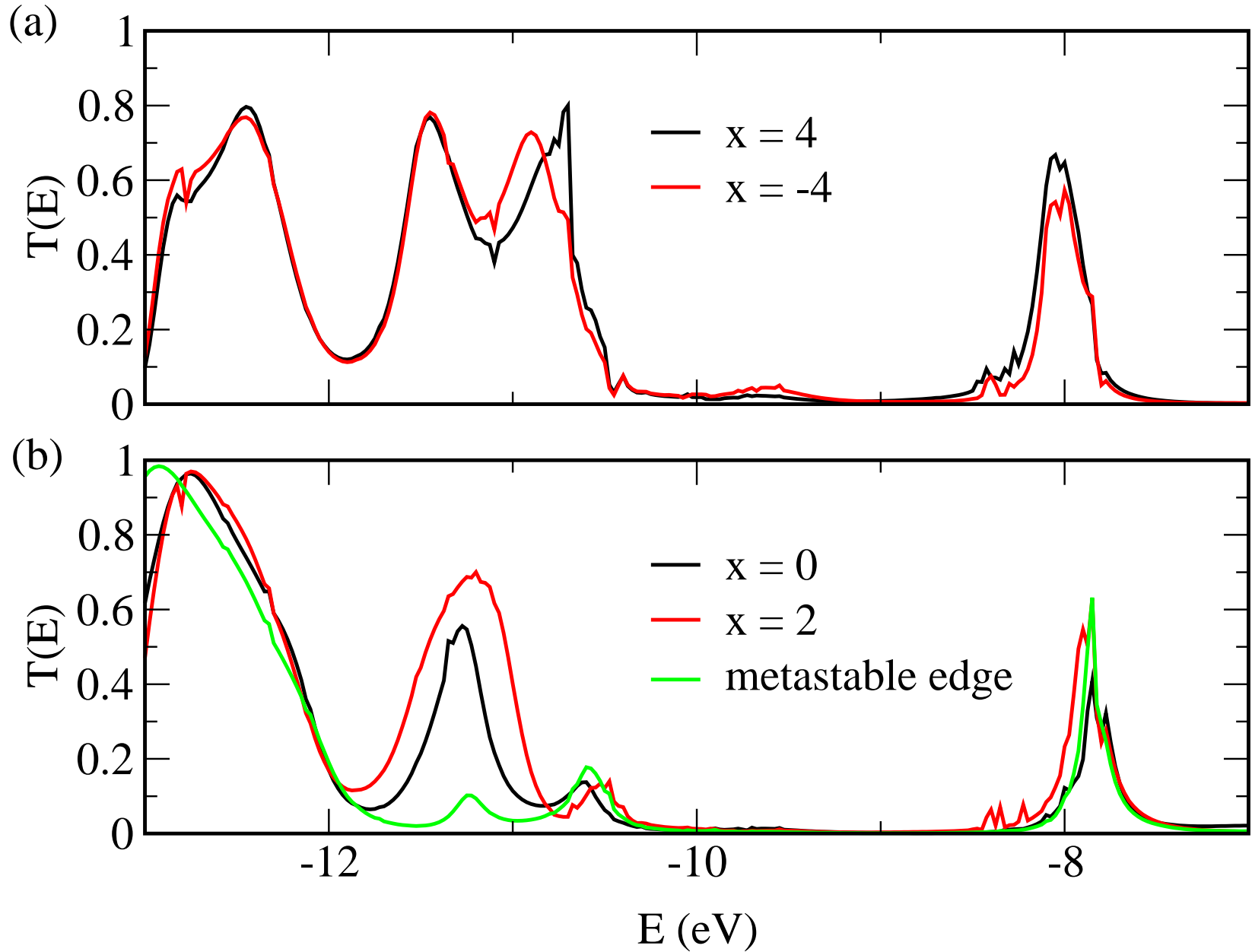
\end{document}